\documentclass[10pt]{article}
\usepackage[final]{graphics}
\usepackage{amsmath,pstricks}
\usepackage{amsfonts,amsbsy}

\font\tenimbf=cmmib10 at 10pt
\font\sevenimbf=cmmib10 at 7pt
\font\fiveimbf=cmmib10 at 5pt
\newfam\imbf
\textfont\imbf=\tenimbf
\scriptfont\imbf=\sevenimbf
\scriptscriptfont\imbf=\fiveimbf

\def\empile#1\over#2{\mathrel{\mathop{\kern 0pt#1}\limits_{#2}}}

\def\bs{\boldsymbol}

\def\k{{\boldsymbol k}}
\def\x{{\boldsymbol x}}
\def\y{{\boldsymbol y}}

\begin{document}
\title{\bf Effective potential in the BET formalism}
\author{A. Bessa$^{(1)}$, C.A.A. de Carvalho$^{(2)}$, 
E.S. Fraga$^{(2)}$, F. Gelis$^{(3)}$}
\maketitle
\begin{center}
\begin{enumerate}

\item Escola de Ci\^encias e Tecnologia\\
 Universidade Federal do Rio Grande do Norte\\
Caixa Postal 1524, Natal, RN 59072-970, Brazil \\
\item Instituto de F\'\i sica\\
Universidade Federal do Rio de Janeiro\\
C.P. 68528, Rio de Janeiro, RJ 21941-972, Brazil
\item Institut de Physique Th\'eorique\\
CEA/DSM/Saclay, Orme des Merisiers\\
91191 Gif-sur-Yvette cedex, France
\end{enumerate}
\end{center}

\begin{abstract}
We calculate the one-loop effective potential at finite
temperature for a system of massless scalar fields with quartic
interaction $\lambda\phi^4$ in the framework of the {\it boundary effective theory} (BET) 
formalism. The calculation relies on the
solution of the classical equation of motion for the field, and
Gaussian fluctuations around it. Our result is non-perturbative and 
differs from the standard one-loop effective potential for field
values larger than $T/\sqrt{\lambda}$.
\end{abstract}

\maketitle

\section{Introduction}\label{sec:intro}

In thermal systems containing massless bosons, a direct implementation of perturbation 
theory for the calculation of thermodynamic quantities is problematic. Vanishing Matsubara 
modes bring, then, infrared divergences that render the naive perturbative series essentially 
non-convergent and meaningless \cite{FTFT-books}. Therefore, the only way to proceed with 
a sensible perturbative calculation in this realm is to reorganize the diagrammatic series by 
resumming certain classes of diagrams. In particular, in the early days of the computation of 
the one-loop effective potential, it was immediately realized that one should resum the so-called 
ring diagrams even in the limit of weak coupling, and that this recipe provides a thermal mass for 
the bosonic (thermal) propagator \cite{Dolan:1973qd}. Resummations of this sort can be performed 
and improved in different ways. We refer the reader to the reviews of Refs. 
\cite{Blaizot:2003tw,Kraemmer:2003gd,Andersen:2004fp} and to Ref. \cite{Bessa:2007vq} for 
a longer discussion and a list of more specific references. 

In a recent paper \cite{Bessa:2010tp}, we have proposed an alternative approach to thermal field 
theories, which we denoted boundary effective theory (BET). In the present paper, we apply the BET formalism 
to compute the one-loop effective potential at finite temperature for a system of massless scalar 
fields with quartic interaction $\lambda\phi^4/4!$. The calculation relies on the solution of the classical 
equation of motion for the field, and Gaussian fluctuations around it. Our result is non-perturbative 
and differs from the standard one-loop effective potential \cite{Dolan:1973qd} for field values larger 
than $T/\sqrt{\lambda}$.

The boundary effective theory (BET) is a natural way to organize the
calculation of the partition function of a quantum system at finite
temperature, where one slices the functional space of $\beta$-periodic
fields into sectors where the boundary value of the field is fixed \cite{Bessa:2010tp}.
More precisely, this amounts to writing the partition function $Z$ as
\begin{equation}\label{eq:Z}
Z=\int[D\phi_0(\x)]\;\rho_\beta[\phi_0(\x),\phi_0(\x)]\;,
\end{equation}
where $\phi_0(\x)$ is the field at the boundaries of the imaginary
time interval, and $\rho_\beta[\phi_0(\x),\phi_0(\x)]$ is the functional density matrix
diagonal element, given by the
functional integration over all the fields $\phi(\tau,\x)$ that have
this particular value at the time boundary,
\begin{equation}
\rho_\beta[\phi_0(\x),\phi_0(\x)] \equiv
 \int\limits_{\phi(0,\x)=\phi(\beta,\x)=\phi_0(\x)}[D\phi(\tau,\x)]\;
\;\;\;\; e^{-S_{_E}[\phi]}\; ,
\end{equation}
where $S_{_E}[\phi]$ is the Euclidean classical action. 

The functional density matrix diagonal element $\rho_\beta[\phi_0(\x),\phi_0(\x)]$ is an expression involving only the
static boundary field $\phi_0(\x)$, which already contains all the
temperature dependence. Thus, correlations calculated within this reduced
theory encode all information about the thermal distribution of the
fields $\phi_0(\x)$. 
One can define a dimensionally-reduced action $S_d[\phi_0(\x)]$ through
\begin{align}
\rho_\beta[\phi_0(\x),\phi_0(\x)] = e^{-S_d[\phi_0(\x)]}\;.
\end{align}

The reduced theory also contains the relevant infrared
physics. Indeed, the double integral structure of the partition
function naturally separates the static modes $\phi_0(\x)$. As a
consequence, the reduced theory amounts to a resummation of an
infinite class of diagrams of naive perturbation theory. This was
verified in Ref. \cite{Bessa:2010tj}, where the effective action
obtained in \cite{Bessa:2010tp} was used to compute the
pressure to lowest order in the BET formalism. The resummation of
ring-like diagrams emerges directly from the theory, and the
corresponding pressure is in good agreement with recent calculations
using weak-coupling and screening perturbation 
theory \cite{Andersen:2009ct,Andersen:2000yj,Andersen:2008bz}.

One should notice that BET is not a particular case of a procedure known in the literature as Dimensional 
Reduction (DR)
 (see Refs. \cite{Braaten:1995cm,DR-2,DR-3,DR-4,DR-5,DR-6}). Indeed, DR methods produce
effective theories for the zero Matsubara mode and, as such, 
are high-temperature approximations in character. 
 The procedure that we utilize (BET) yields an {\it alternative} dimensionally-reduced effective theory 
for the physical field $\phi_0(\x)$, and it is essentially different from DR.

The aforementioned calculations using the BET formalism are
non-perturbative, the results being expressed in terms of the
classical solutions $\phi_c$ of the Euler-Lagrange equation. When
these classical solutions are computed exactly, instead of being
expanded perturbatively, they automatically resum the infinite series
of tree diagrams in the strong field regime. Note that the
Euler-Lagrange equations must be solved with non-trivial boundary
conditions at the endpoints of the imaginary time interval:
$\phi_c(\tau,\x)$ must be equal to $\phi_0(\x)$ at the time
boundary. To emphasize this functional dependence, we will denote the
classical solution by $\phi_c[\phi_0(\x)]$. In Ref. \cite{Bessa:2010tp}, it is shown
that the effective action for the fields $\phi_0(\x)$ admits a simple
expression in terms of the classical solution $\phi_c[\phi_0(\x)]$. In
order to extract the explicit dependence on $\phi_0(\x)$, one must
solve the Euler-Lagrange equation for arbitrary boundary conditions,
which is not a feasible task in an interacting theory (but can, in
principle, be done numerically).

In this paper, we solve the full classical equation with constant
boundary conditions $\phi_0(\x)\equiv\phi_0$, allowing for the
calculation of the one-loop effective potential of the theory. The
complicated non-linear dependence of $\phi_c$ on the boundary field
$\phi_0$ leads to a non-perturbative result that should be a good
approximation even at large $\phi_0$.

Finally, it should be clear that the use of the word \lq\lq boundary\rq\rq\,
has no relation to other uses, often in a topological sense, 
such as in holographic gauge-gravity duality, etc. 
 
The structure of the paper is as follows: in Section
\ref{sec:VfromJHEP}, we revisit the usual prescription to obtain the
effective action and the effective potential in the functional integral
formalism; using the results of Ref. \cite{Bessa:2007vq},
we write the effective potential in terms of the solution of a certain
differential equation; in Section \ref{sec:massless}, we apply the
method to the massless $\lambda \phi^4/4!$ theory, and discuss the
results for the effective potential; finally, in
Section \ref{sec:conclusions}, we present our conclusions.

\section{The one-loop effective potential in BET}\label{sec:VfromJHEP}

Let us consider the following classical Euclidean action,
\begin{equation}\label{eq:euclideanaction}
S_{_E}[\phi]=
\int\limits_{0}^{\beta} (d^4x)_{_E}\, \left[\frac{1}{2}\partial_\mu \phi \partial^\mu\phi +  U(\phi)\right]\;,
\end{equation}
where $(d^4x)_{_E}$ is a shorthand for $d\tau\,d^3x$. We assume that
$U(\phi)$ is some single-well interaction potential. Following the
standard procedure for obtaining the effective action, we couple the
boundary field to an external current $j(\x)$, and define the
generating functional for the {\it reduced theory} \cite{itzykson}:
\begin{equation}\label{eq:Z2}
Z[j(\x)]=\int[D\phi_0(\x)]\; e^{-S_j[\phi_0(\x)]}\; ,
\end{equation}
where
\begin{equation}
S_j[\phi_0(\x)] \equiv S_{d}[\phi_0(\x)] -\beta\, \int d^3x\;j(\x)\phi_0(\x)
\end{equation} 
is the action for boundary fields in the presence of $j(\x)$.  The free energy functional
can be obtained from the generating functional $Z[j(\x)]$ as
\begin{equation}
F[j(\x)] = -\frac{1}{\beta}\,\lim_{V \rightarrow \infty}\,\ln Z[j(\x)]\;. 
\end{equation}
By performing a Legendre transform of $F[j(\x)]$, one formally obtains the
effective action
\begin{equation}
\Gamma[\langle\phi_0(\x)\rangle_j] = F[j(\x)] + \int d^3x \;j(\x)\langle\phi_0(\x)\rangle_j\;.
\end{equation}
Its argument is the expectation value of the field in the presence of
the external current. The index $j$ in $\langle \phi_0(\x)\rangle_j$ is to
stress the dependence of the expectation value on the external
current. One can think of $\Gamma$ as {\it minus} the pressure of the
system in response to an external current $j(\x)$. It can be shown
that, at one-loop order, we have $F[j(\x)] = \Gamma[\langle
  \phi_0(\x)\rangle_j]$, and that $\langle \phi_0(\x)\rangle_j$ is the
saddle-point of $S_j[\phi_0(\x)]$.

\begin{figure}[htbp]
\begin{center}
\resizebox*{9cm}{!}{\includegraphics{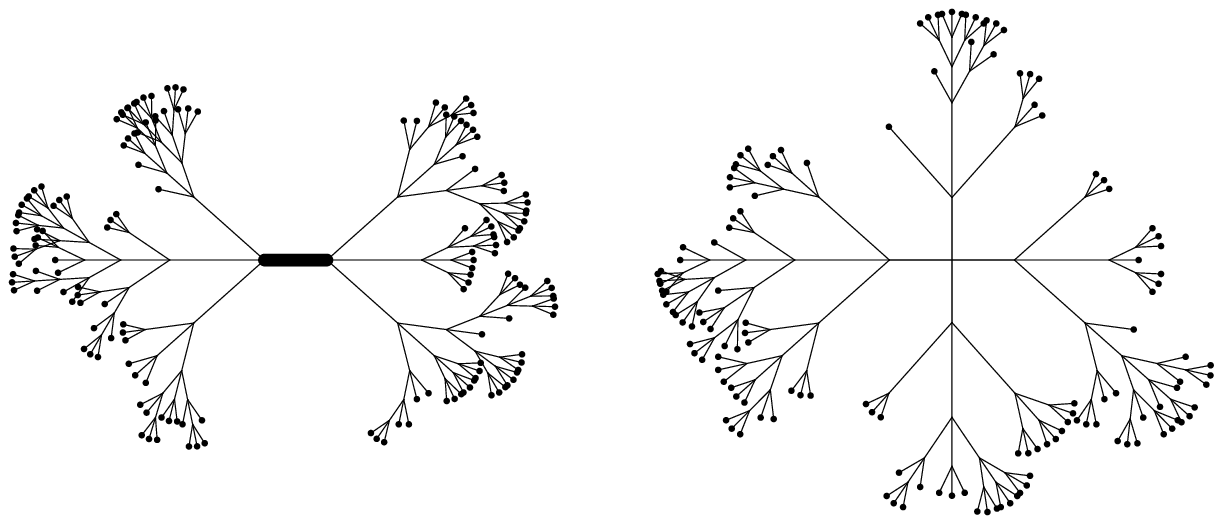}}
\hfil
\resizebox*{4cm}{!}{\includegraphics{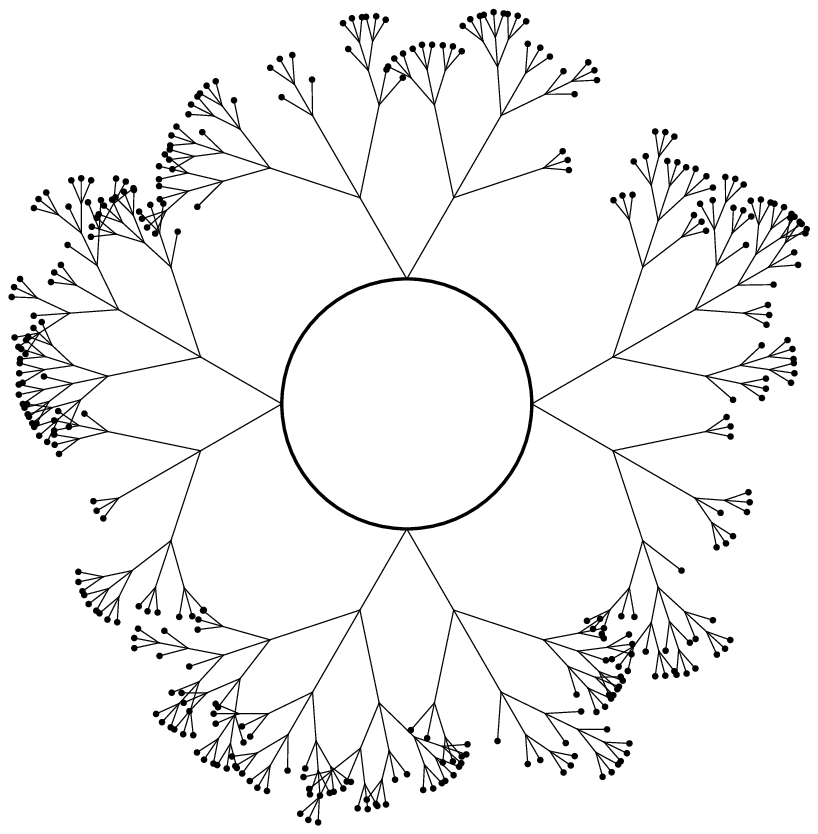}}
\end{center}
\caption{\label{fig:tree}Diagrammatic representation of the content of
  eq.~(\ref{Gammafinal}). The black dots at the endpoints of the trees
  represent the boundary field $\phi_0(\x)$ at which the effective
  action is evaluated.}
\end{figure}

In Ref. \cite{Bessa:2010tp}, it was shown that the one-loop effective action
for ${\phi}_0(\x)$ is given by
\begin{equation}\label{Gammafinal}
\beta\,\Gamma[\phi_0(\x)] = S_{_E}[\phi_c[\phi_0(\x)]] 
+\frac{1}{2} {\rm Tr}\,\ln 
\left (\Delta_F^{-1} + U^{\prime\prime}(\phi_c[\phi_0(\x)]) \right )\;,
\end{equation}
where $\Delta_F$ is the time-ordered thermal
propagator and $\phi_c[\phi_0(\x)](\tau,\x)$ is the classical solution of
the Euler-Lagrange equation for fixed time boundary value
$\phi_0(\x)$, i.e.
\begin{eqnarray}\label{eq:classiceqforphi0}
  &&\square_{_E}\phi_c(\tau,\x)+U'\left(\phi_c(\tau,\x)\right)=0
  \; ,\nonumber\\
  && \phi_c(0,\x)=\phi_c(\beta,\x)=\phi_0(\x)\;,
\label{eq:EOMc}
\end{eqnarray}
with $\square_{_E}\equiv -(\partial_\tau^2+{\bs\nabla}^2)$ the
Euclidean D'Alembertian operator. We see that, when written in terms
of $\phi_c[\phi_0(\x)]$, $\Gamma[\phi_0(\x)]$ is the same functional
as the one-loop effective action at zero-temperature.  In terms of
graphs, eq.~(\ref{Gammafinal}) includes the diagrams represented in the
figure \ref{fig:tree}.
In this figure, the first two terms represent the classical action
evaluated at the classical field configuration $\phi_c$. The tree
structure of $\phi_c$ when expressed in terms of $\phi_0(\x)$ is
manifest in the Green's formula that relates $\phi_c$ to its
boundary value $\phi_0$,
\begin{equation}
\phi_c(\tau,\x)
+\int d\tau^\prime d^3\y\;
G(\tau,\x;\tau^\prime,\y)\,U^\prime(\phi_c(\tau^\prime,\y))
=
\int d^3\y\;\phi_0(\y)\,
\Big[
\partial_{\tau^\prime}G(\tau,\x;\tau^\prime,\y)
\Big]_{\tau^\prime=0}^{\tau^\prime=\beta}\; ,
\label{eq:green}
\end{equation}
where $G(\tau,\x;\tau^\prime,\y)$ is the Green's function of the
Euclidean D'Alembertian defined by
\begin{equation}
\square_{_{E}}^y G(\tau,\x;\tau^\prime,\y)=\delta(\tau-\tau^\prime)\delta(\x-\y)\;,
\qquad
G(\tau,\x;0,\y)=G(\tau,\x;\beta,\y)=0\; .
\end{equation}
The last term in the figure \ref{fig:tree} is the diagrammatic content
of the term $\frac{1}{2} {\rm Tr}\,\ln(\cdots)$ in
eq.~(\ref{Gammafinal}). Obviously, the non-linearities in $\phi_0$ are
important only at large $\phi_0$ -- when $\phi_0$ is small (compared
to the temperature), all the trees in the figure \ref{fig:tree}
simplify into their lowest order term in $\phi_0$ (which is linear in
$\phi_0$ and amounts to solving eqs.~(\ref{eq:EOMc}) by neglecting the
non-linear potential $U^\prime(\phi_c)$).

In order to obtain the explicit dependence of $\Gamma[\phi_0(\x)]$ on
$\phi_0(\x)$, we must solve Eq. \eqref{eq:classiceqforphi0} for
arbitrary boundary conditions, which is in general unfeasible for
boundary fields with an arbitrary $\x$ dependence. A simpler, yet very useful quantity
to compute, is the effective potential, which is essentially the
effective action evaluated for uniform boundary field configurations,
\begin{equation}\label{def:Veff}
V_{\rm eff}(\phi_0) 
=
(\Gamma[\phi_0] - \Gamma[0])/V \qquad\mbox{(for constant } \phi_0(\x)\equiv\phi_0\hbox{)}  ,
\end{equation}
where $V$ is the volume. In this case, the classical solution
$\phi_c[\phi_0]$ depends only on $\tau$. Definition \eqref{def:Veff} is such that $V_{\rm eff}(0)=0$. 
As shown in Ref. \cite{Bessa:2010tp}, the quantity $\Gamma[0]/V$ is the negative of the free pressure, 
i.e. $(-\pi^2T^4/90)$.

The calculation of the term $\frac{1}{2}{\rm Tr}\,\ln(\cdots)$ in Eq. (\ref{Gammafinal}) was done
in Ref. \cite{Bessa:2007vq} as an intermediate step to obtain the
pressure in the context of a semiclassical approximation. There, it was
shown that this quantity can be expressed, for constant $\phi_0$, in
terms of solutions of the equation for small field perturbations
propagating on top of the classical solution $\phi_c$
\begin{eqnarray}
 \frac{1}{2}\; {\rm Tr}\,\ln \left (\Delta_F^{-1} + U^{\prime\prime}(\phi_c[\phi_0]) \right )
 =
\frac{V}{2}
\int^{\Lambda}\frac{d^3\k}{(2\pi)^3}
\ln \left[ 2\left ({\eta}(\beta,\k^2)-1 \right )\right]\;,\label{eq:final}
\end{eqnarray}
where  $\eta(\tau;\k^2)$ is the solution of
\begin{subequations}\label{eq:odegeral}
\begin{gather}
\left[\partial_{\tau}^2 - k^2 -U''(\phi_c[\phi_0](\tau))\; \right]\eta(\tau,\k^2)=0 \;,\\
\eta(0;\k^2)=1\; ,\;\;\;\frac{d\eta}{d\tau}(0;\k^2)=0\;.
\end{gather}
\end{subequations}
Therefore, we obtain the following (non-renormalized) expression for
the effective potential,
\begin{equation}\label{eq:effV_nR}
\beta V\,V_{\rm eff}(\phi_0) =  S_{_E}[\phi_c[\phi_0]] + \frac{V}{2}
\int^{\Lambda}\frac{d^3\k}{(2\pi)^3}
\ln \left[ 2\left ({\eta}(\beta,\k^2)-1 \right )\right]\;-\beta\Gamma[0]\;,
\end{equation}
where the integral over 3-momenta is regularized by the introduction
of a cut-off $\Lambda$. 

In order to renormalize \eqref{eq:effV_nR}, we add the standard
one-loop counterterms ${\rm C.T.}$ (obtained by using the same cutoff
regularization as the one used in Eq.~(\ref{eq:effV_nR})), and subtract
the zero-point energy term $\beta k/2$ from the classical action,
obtaining
\begin{equation}\label{eq:effV_R}
\beta V\,V_{\rm eff}{}_R(\phi_0) =  S_{_E}[\phi_c[\phi_0]]
+ \lim_{\Lambda\rightarrow \infty}
\frac{V}{2}\int^{\Lambda}\frac{d^3\k}{(2\pi)^3}
\Big[
\ln \left[ 2\left ({\eta}(\beta,\k^2)-1 \right )\right] - \beta k
\Big]
-{\rm C.T.} - \beta\Gamma[0]\;,
\end{equation}
where the counterterms read
\begin{align}
{\rm C.T.} = V\,\frac{C_1}{2}\int d\tau\,\phi_c^2(\tau) + V\,\frac{C_2}{4}\int d\tau\,\phi_c^4(\tau)\;,
\end{align}
with
\begin{align}\label{CT}
C_1 \equiv \frac{\lambda}{2}\int^{\Lambda}\frac{d^4k}{(2\pi)^4}\Delta_F^0(k) = \frac{\lambda}{16\pi^2}\,\Lambda^2
\end{align}
and
\begin{align}\label{CT2}
C_2 \equiv -\frac{\lambda^2}{4}\int^{\Lambda} \frac{d^4k}{(2\pi)^4}\Delta_F^0(k)\Delta_F^0(k+\mu) = - \frac{\lambda^2}{32\pi^2}\,\left (\ln\frac{\Lambda}{\mu} + \frac{1}{2} \right)\;,
\end{align}
where $\mu$ is the renormalization scale.

\section{Results for a massless theory with quartic interaction}
\label{sec:massless}

When $U(\phi) \equiv \lambda \,\phi^4/4!$, the classical equation of motion is
\begin{eqnarray}
&&-\partial_\tau^2\phi_c(\tau)+\frac{\lambda}{6}\phi_c^3(\tau)=0\; ,
\nonumber\\
&&\phi_c(0)=\phi_c(\beta)=\phi_0\; .
\end{eqnarray}
The solution of this equation with the required boundary condition is given by
\begin{equation}\label{eq:phi0quartic}
\phi_c(\tau)=
\sqrt{\frac{6}{\lambda}}\,\varphi_t \;{\rm nc}\left(\varphi_t(\tau-\beta/2),1/\sqrt{2}\right)\; ,
\end{equation}
where ${\rm nc}$ is one of the twelve Jacobi Elliptic Functions \cite{AS}, and
$\varphi_t$ is defined implicitly by the following equation:
\begin{eqnarray}
\phi_0=\sqrt{\frac{6}{\lambda}}\,\varphi_t 
\;{\rm nc}(\varphi_t\beta/2,1/\sqrt{2})\;.
\end{eqnarray}

\begin{figure}[htb!]
\begin{center}
\rotatebox{-90}{%
     \resizebox{9cm}{!}{\includegraphics{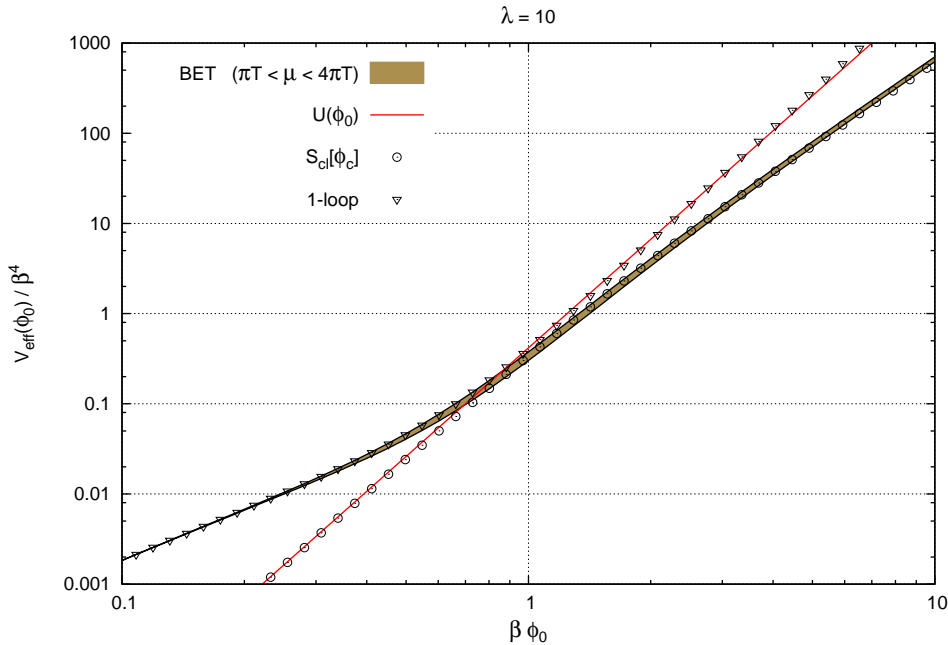}}}
\end{center}
\caption{\label{fig:effV10}Plot of $\beta^4 V_{\rm eff}(\phi_0)$ for
  a massless quartic interaction with $\lambda = 10$. The shaded band
  represents the BET result, with a renormalization scale varying in
  the range $\mu\in[\pi T,4\pi T]$. The thin solid line is the
  classical potential, i.e. $U(\phi_0) = \lambda\phi_0^4/4!$ in this case. The
  open circles represent the effective potential obtained from the
  classical action evaluated at the solution $\phi_c$ of the
  Euler-Lagrange equation with boundary value $\phi_0$. The open
  triangles represent the standard 1-loop result, evaluated at
  $\mu=2\pi T$.}
\end{figure}

Substituting \eqref{eq:phi0quartic} in Eq. \eqref{eq:odegeral}, we obtain
\begin{subequations}\label{eq:odephi4}
\begin{gather}
\left[\partial_{\tau}^2-\k^2
-3\,\varphi_t^2\,{\rm nc}^2\left(\varphi_t(\tau-\beta/2),1/\sqrt{2}\right)\right]\eta(\tau,\k^2)=0\,,
\;\\
\eta(0;\k^2)=1\; ,\qquad\frac{d\eta}{d\tau}(\beta;\k^2)=0\;.
\end{gather}
\end{subequations}
We solve Eq. \eqref{eq:odephi4} numerically, and use $\eta(\tau,\k^2)$
 in Eq. \eqref{eq:effV_R} in order to obtain the
one-loop effective potential in the BET approach.

\begin{figure}[htb!]
\begin{center}
\rotatebox{-90}{%
     \resizebox{9cm}{!}{\includegraphics{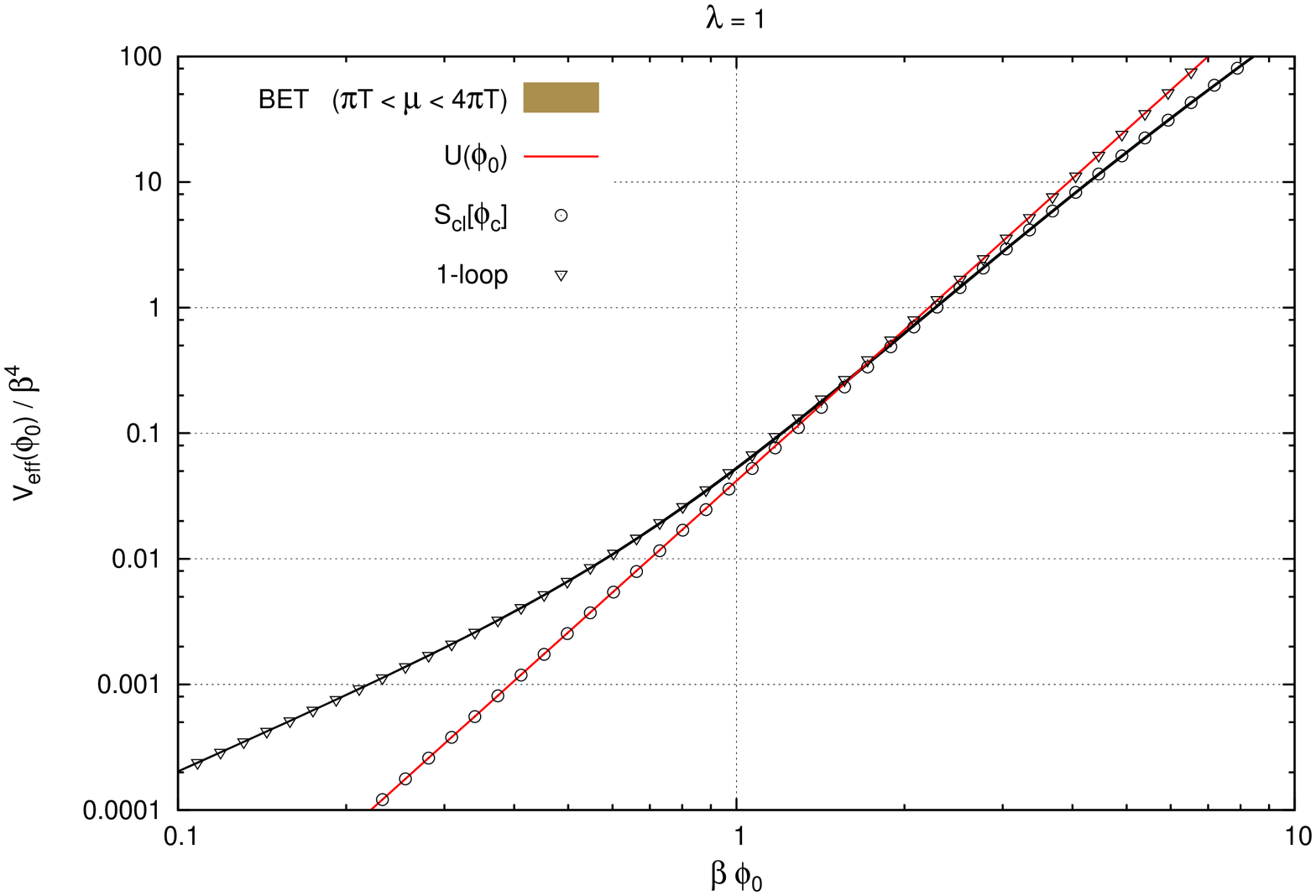}}}
\end{center}
\caption{\label{fig:effV1}Plot of $V_{\rm eff}(\phi_0)/\beta^4$ for
  a massless quartic interaction with $\lambda = 1$. Definitions are the 
  same as in Fig. \ref{fig:effV10}.}
\end{figure}

The BET effective potential (in units of $T^4$) is plotted in
Fig. \ref{fig:effV10} as a function of the dimensionless field
$\beta\phi_0$ for a coupling constant $\lambda = 10$ (the results
for $\lambda=1$ are exhibited in Fig. \ref{fig:effV1}). The
result is displayed in the form of a band corresponding to a
variation of the renormalization scale within the interval $\pi T<\mu<4\pi T$. One can see
that the residual sensitivity to the renormalization scale is fairly
small, suggesting that higher-order corrections are well under
control. In Fig. \ref{fig:effV1}, which shows the results for
$\lambda=1$, this band is so narrow that it appears as a line.

Our result is first compared to the classical potential itself,
$U(\phi_0)$. One can see that it differs from the classical
potential both at small field, due to the appearance of a quadratic
mass term, and at large field due to large non-linear corrections.  At
small field, the standard one-loop result and BET are in good agreement,
since both incorporate the effect of the thermal mass $m^2 =
\lambda T^{2}/24$.  However, for larger values of the field, differences
appear. In fact, the one-loop result has an asymptotic behavior close
to $(\lambda/24)(\beta\phi_0)^4$ (the one-loop correction becomes
very small at large field, and one simply recovers the classical
potential), while the BET result approaches
$\sqrt{\lambda/27}\,(\beta\phi_0)^3$. One can see that this result is
in fact dictated by the behavior of the classical solution
$\phi_c[\phi_0]$: when the boundary field $\phi_0(\x)=\phi_0$ is
large, the non-linear term in the classical Euler-Lagrange equation is
very important, and the solution $\phi_c$ becomes a strongly
non-linear function of the boundary field value $\phi_0$. This
non-linearity alters significantly the behavior of the BET effective
potential at large $\phi_0$. In fact, one can see that in this
regime the BET effective potential is well approximated by the
classical action evaluated at the solution $\phi_c$. This suggests
that for large fields, the main effect comes from the infinite sum of
the non-linear tree level contributions.

\section{Conclusions}
\label{sec:conclusions}

The boundary effective theory (BET) framework provides a way to control the infrared divergences 
of thermal field theory in a well-defined and relatively simple way \cite{Bessa:2010tp}. Previously, 
we had computed the pressure of a massless hot scalar $\lambda\phi^{4}$ 
theory \cite{Bessa:2010tj}, obtaining excellent agreement with up-to-date results from weak-coupling 
and screening perturbation theory \cite{Andersen:2009ct,Andersen:2000yj,Andersen:2008bz}. 
In this paper, we have applied this method to the computation of the one-loop effective potential, 
following our previous work on the semiclassical thermodynamics of scalar fields \cite{Bessa:2007vq}.

The effective potential obtained within the BET formalism perfectly reproduces the standard one-loop 
result \cite{Dolan:1973qd} for small fields, as expected, since the BET effective action contains 
very naturally the effect of the thermal mass. For large fields, BET goes beyond by incorporating the 
nonlinear corrections that become more and more important and are not captured by the standard 
one-loop calculation. We have also shown that our results are very stable with respect to variations 
of the renormalization scale, signaling a good behavior of the (semiclassical) series.

A natural, and very useful, extension of this work would be treating the case of thermal symmetry 
restoration in the case of a double-well classical potential for the scalar field, with its consequences 
for spontaneous symmetry breaking and the description of phase transitions. However, this is not a 
straightforward extension, since the case of multiple wells presents nontrivial features related to the appearance 
of caustics and complex trajectories in the calculation of the semiclassical density matrix (see 
Refs. \cite{deCarvalho:1998ff,deCarvalho:2001vk} for a discussion). Nevertheless, we believe that 
the nonlinear corrections captured by the BET approach can be of great relevance in the description 
of phase transitions. Results in this direction will be reported in the future \cite{future}.

\section*{Acknowledgments}
The authors thank the support of CAPES-COFECUB, project $663/10$. 
The work of A.B., C.A.C. and E.S.F. was partially supported by 
CAPES, CNPq, FAPERJ and FUJB/UFRJ.


\end{document}